# Phase change-photonic framework for terahertz wave control

Prakash Pitchappa[1,2], Abhishek Kumar[1,2], Saurav Prakash[3,4], Hariom Jani[3,4], Thirumalai Venkatesan[3,4,5,6,7] and Ranjan Singh[1,2]*

[1]Division of Physics and Applied Physics, School of Physical and Mathematical Sciences, Nanyang Technological University, 21 Nanyang Link, Singapore-637371

[2]Centre for Disruptive Photonic Technologies, The Photonic Institute, 50 Nanyang Avenue, Singapore-639798

[3]NUSNNI-NanoCore, National University of Singapore, Singapore

[4]NUS Graduate School for Integrative Science and Engineering, National University of Singapore, Singapore 117456

[5]Department of Physics, National University of Singapore, Singapore 117542

[6]Department of Electrical and Computer Engineering, National University of Singapore, Singapore 117583

[7]Department of Materials Science and Engineering, National University of Singapore, Singapore 117575

*Correspondence author E-mail: ranjans@ntu.edu.sg

The advancement in capabilities enabled by handheld devices and the prospective internet of things has led to an explosion of wireless data consumption in the last decade. The ever-increasing demand for high speed data transfer has projected the unprecedented importance of terahertz spectral range for next generation wireless communication systems. However, versatile interaction and active manipulation of terahertz waves required for a wide-range of terahertz technologies are either unavailable or unscalable. Here, we report on an integrated platform of chalcogenide phase change material, germanium antimony telluride (GST), with metamaterials for multidimensional manipulation of terahertz waves. The thermal, electrical and optical stimulation of GST provide multilevel non-volatility, sub-wavelength spatial selectivity and variable ultrafast switching of terahertz waves, respectively. These multifaceted response of phase change material can be used in tandem with metamaterial that enables strong field confinement along with on-demand spectral response and frequency selectivity. The unique and multifarious properties of GST combined with the versatility and scalability of metamaterial design and ease of fabrication could aid in the development of advanced devices operating in terahertz range, such as spatio-temporal light modulators, neuromorphic photonic devices, compressive imagers, and switchable holograms, vital for the next generation wireless communication systems.

The terahertz part of the electromagnetic spectrum possesses numerous promising applications such as high-speed wireless communication, security screening, chemical identification and non-destructive biosensing.[1] However, the terahertz spectral region has remained technologically uncharted, due to the lack of efficient devices to generate, manipulate and detect terahertz waves.[2] In the past decade, there has been a significant progress in active control of terahertz waves using metamaterials integrated with active materials and are popularly termed as "terahertz metadevices".[3-6] Metamaterial is an array of subwavelength sized resonators whose electromagnetic properties are primarily determined by the resonator geometry, which makes them functionally rich and spectrally scalable. Furthermore, metamaterial resonators provide strong confinement of electromagnetic fields in a substantial subwavelength region providing enhanced light-matter interaction. This has led to the development of energy efficient terahertz switches,[7] spectrally tunable filters,[5] efficient polarizers,[8] beam steerers[9] and high-performance sensors[10]. A wide array of active materials are explored for the realization of terahertz metadevices including semiconductors (Si, Ge, GaAs, perovskites),[3, 11-13] liquid crystals,[14] 2D materials (graphene, $MoS_2$),[15-17] superconductors,[18-21] and phase change materials ($VO_2$, GeTe).[22-25] However, most of these approaches are limited to a single functionality in terms of switching between binary states, fixed modulation speed, and volatile response. Furthermore, these active materials are usually responsive to a single driving field, one of electrical, thermal, or optical stimulus, which seriously hinders the applicability of metadevices for multidimensional manipulation of terahertz waves. Phase change materials with contrasting material properties at different crystallographic phases would be an ideal material platform for the realization of multifunctional terahertz metadevices.[26] The most popular choice of phase change materials for terahertz metadevices has been vanadium dioxide ($VO_2$), which has been exploited for varied functionalities enabled by its sharp metal to insulator phase transition.[22, 24, 27, 28]

However, the practical applicability of VO$_2$ for terahertz metadevices are hindered by its limited multilevel response, non-volatility at elevated temperature, predefined singular ultrafast optical response and complexity of fabrication.[26]

Alternatively, chalcogenide phase change materials undergo phase change mediated by nucleation dynamics.[29] Hence, by continuously varying the proportion of crystalline phases in the amorphous matrix, we can achieve analog response rather than binary switching states. More importantly, the analog states are non-volatile with zero hold power.[30] Preliminary study on chalcogenide phase change material, germanium telluride (GeTe), for active terahertz control with non-volatility has been reported.[25] However, the versatile nature of the chalcogenide phase change materials and their potential in multidimensional terahertz wave manipulation remains largely unexplored. Traditionally, Germanium antimony telluride (GST) was the popular choice of chalcogenide phase change material that fuelled the optical memory era, owing to the best combination of higher switching speed and longer state retention time.[31] In more recent times, GST has led to the demonstration of various hallmark nanophotonic devices, such as integrated all-photonic non-volatile memory,[32] optical colour rendering,[33] nanopixel displays[34] and reconfigurable nanoplasmonics devices.[35, 36] The applicability of GST for various photonic technologies comes from its three crystallographic phases - amorphous, metastable FCC (face centred cubic) and stable HCP (hexagonal close packed) lattice structures, which possess strikingly different electronic and optical properties. However, the exploration of GST has been largely limited to visible and infrared frequencies.[37]

Here, we report GST integrated with resonant metamaterial devices for multidimensional control of terahertz waves. The versatile and unique properties of GST allow for the realization of various terahertz functionalities such as thermally switched multilevel non-volatile states, electrically controlled spatial light modulation and optically controlled ultrafast resonance switching. These unique features could be utilized in conjunction with each other and would enable novel devices for advanced manipulation of terahertz waves. Furthermore, the simplicity of the device fabrication allows for the translation of these multifunctional metadevices on to flexible substrates, thereby opening new avenues for potential applications.

GST is a ternary compound made of varying proportions of germanium (Ge), antimony (Sb) and tellurium (Te). Here, we use $Ge_2Sb_2Te_5$, which is the most widely used stoichiometry of GST for photonic devices. Thin film of GST was sputter deposited from $Ge_2Sb_2Te_5$ stoichiometric target on to quartz substrate (see Methods). Raman spectroscopy was used to characterize the phase of the sputter deposited GST film (see Methods). The as-deposited GST film is amorphous as shown in **Figure 1a** (see Supplementary Figure 1). The crystallization temperature of GST film to FCC phase has been reported as ~ 150 °C and to HCP phase as ~ 250 °C.[38] For our experiments, GST samples were annealed at temperatures of 180 °C and 260 °C for 60 mins in ambient conditions. The inspection of the film under optical microscope after annealing process showed no signs of film degradation (see insets of Supplementary Figure 2b and 2c). However, when the annealing was increased to 325 °C, light and dark spots were observed indicating lateral inhomogeneity in the GST film (see inset of Supplementary Figure 2d). Hence, we have refrained from using very high temperature (> 260 °C) annealing conditions. Raman spectra for the GST film before and after 180 °C and 260 °C annealing is shown in **Figure 1a**. Multiple spots over the film surface were characterized using Raman spectroscopy to confirm uniformity and consistent film quality (see Supplementary Figure 2). The detailed analysis of the Raman peaks and their evolution with annealing temperatures were carried out (see Supplementary Figure 3 and Supplementary Table 1). The phase composition of GST after each annealing temperature was elucidated based on the comparison with earlier reported literature (see Supplementary Table 2). The as-deposited GST thin film is amorphous, 180 °C annealing shows dominant FCC peaks with very weak HCP peaks, and 260 °C annealing shows stronger FCC and HCP peaks. The concurrent observation of FCC and HCP peaks in our Raman data from individual spots indicates that a mixed phase coexists at scales smaller than ~1 µm$^2$, which is the spot size of our Raman spectrometer (see Methods). X-ray diffraction was used to confirm the observed phases of GST film after different annealing temperatures (see Supplementary Figure 4). It supports the phase identification performed via Raman Spectroscopy. The terahertz properties of GST film were characterized using terahertz time domain spectroscopy system (see Methods). The terahertz optical constants were extracted from the transmitted terahertz time signal through the GST film with quartz substrate as reference (See Supplementary Figure 5). The extracted terahertz conductivity of as-deposited, 180 °C and 260 °C annealed GST films shows remarkably contrasting values as presented in **Figure 1b** (see Methods). A strong increase in terahertz conductivity is observed with increasing order of crystallinity in GST film, similar to the trend reported for DC conductivity.[39] The terahertz conductivity of as-deposited GST film is ~4 S/cm and increases to ~40 S/cm and ~350 S/cm for 180 °C and 260 °C annealed films, respectively. Further, a more continuous change in terahertz conductivity, i.e. an analog response, can be achieved by finely manipulating the percentage of crystalline sites in the as-deposited GST films through control of annealing time at specific temperature. It is also important to note that these analog states of GST are non-volatile with zero hold power and are expected to have retention time of years.[40]

We exploit the continuously varying terahertz conductivity of GST thin film for functional terahertz metadevices by integrating with resonant metamaterials. Asymmetric split ring resonator (ASRR) is chosen as the metamaterial unit cell. The ASRR is a ring resonator with dual split capacitive gaps on opposite arms, and one of the gaps is displaced from the central position to create a structural asymmetry as shown in **Figure 1c**. When terahertz wave with electric field polarized perpendicular to the gap bearing arms is shone on the ASRR metamaterial at normal incidence, a sharp asymmetric Fano resonance and broad dipole resonance are excited (see Supplementary Figure 6a and 6b). The Fano resonance is identified by its characteristic circulating current configuration in the ASRR, while at dipole resonance, the currents are oriented parallel to the incident electric field in two arms of the ASRR (see insets of Supplementary Figure 6b). The high $Q$ asymmetric Fano resonance in the ASRR arises from the structural asymmetry, which leads to an interference between a sharp discrete resonance and a much broader continuum-like spectrum of dipole resonance. This narrow resonance arises from a sub-radiant dark mode for which the radiation losses are completely suppressed due to the structure's weak coupling to free space. Fano resonance is selected for our study, owing to their high $Q$ response that enables enhanced sensitivity to external perturbations and is quintessential for probing the analog switching states of GST.[13] The optical image of the fabricated hybrid GST-Fano metadevice (GFMD) is shown in **Figure 1c**, along with the geometrical parameters of the Fano resonator in the inset. The GFMDs showed no sign of degradation after annealing (see Supplementary Figure 7). **Figure 1d** shows the terahertz transmission response of GFMD with as-deposited, 180 °C and 260 °C annealed GST film. For the as-deposited GFMD, the Fano and dipole resonances are measured at 0.72 THz and 1.13 THz, respectively. Finite difference time domain modelling was used to elucidate the nature of resonance excitations. The resonance excitation at 0.72 THz and 1.13 THz was confirmed to be Fano and dipole resonances, respectively (see Supplementary Figure 6b). When annealed to 180 °C for 60 mins, the Fano resonance of GFMD was completely switched off, while the dipole resonance was red-shifted with a small modulation of resonance strength. The modulation of the resonances is caused due to the increased conductivity, while the red shift in resonance frequency is due to the increased refractive index of 180 °C annealed GST film (see Supplementary Figures 6c and 6d). The ultrasensitive nature of Fano resonance enables complete modulation even for small increase in the GST conductivity, compared to the highly radiative dipole resonance which shows significantly small modulation. With further annealing to 260 °C, the dipole resonance modulation is increased with a larger red-shift in resonance frequency, which is expected from the corresponding increase in terahertz conductivity and refractive index for 260 °C annealed GST. Furthermore, the analog terahertz response of GST was exploited to demonstrate multilevel resonance switching by controlling the annealing time at 180 °C as shown in **Figure 1e**. The ultrasensitive nature of Fano resonance was used for probing of fine increment in the terahertz conductivity of GST film. The Fano resonance modulation (FMR) was calculated as $T_F$-$T_I$/$T_I$ x 100 %, where $T_I$ and $T_F$ are the Fano resonance strength (peak to peak) before and after annealing at 180 °C for a constant time, respectively. The dependence of Fano resonance modulation on increasing annealing time at 180 °C shown in **Figure 1f,** reveals an exponential dependence. Each of these analog states are non-volatile with retention time of years (see Supplementary Figure 8). It is interesting to note that GFMD not only provides a two-level (ON-OFF) switching through control of annealing temperature, but also allows for a variable multilevel switching by controlling the annealing time at different annealing temperatures.

Electrical control of terahertz response is extremely critical for the realization of miniaturized solutions and spatially selective reconfiguration.[41, 42] The spatial localization aspect of electrical control was exploited to demonstrate 2 x 2 multicolor spatial light modulator (SLM) for terahertz waves. Electrical switching of GST is mediated by Joule heating, which was confirmed by Raman spectroscopy (see Supplementary Figure 9). The signaling path of applied current was isolated to achieve independent control of each quadrant of the GFMD, thereby forming an SLM device. The unit cells of each quadrant were electrically connected through metallic interconnects between two electrodes as shown in **Figure 2a**, while keeping them electrically isolated from the other three quadrants of the SLM. Each quadrant of the SLM (FR1, FR2, FR3, and FR4) was designed to have spectrally shifted Fano and dipole resonance frequencies by varying the unit cell dimension ($l$) as shown in **Figure 2b** (see Supplementary

Figure 6e), to clearly present the spatially selective modulation. The Fano resonance frequency for FR1, FR2, FR3 and FR4 with as-deposited GST film was measured to be 0.69, 0.64, 0.6 and 0.56 THz, respectively as shown in **Figure 2c**. The spatially selective control is demonstrated by independently switching the Fano resonance of FR1 by passing a constant current through its metallic interconnect for 15s. The Fano resonance remained unchanged until the current was increased to 700 mA at which point, Fano resonance was modulated by ~25 %. The modulation was further increased to ~75% for 800 mA and was completely modulated at 850 mA (see Supplementary Figure 10a). We could readily achieve finer levels of resonance modulation by controlling the applied current and/or hold time. When FR1 was switched, the resonances of other parts of SLM (FR2, FR3 and FR4) remained unchanged as shown in **Figure 2c**. Following the initial demonstration of spatial selective switching, the sequential switching of 2 x 2 SLM with variable multilevel states was implemented (see Supplementary Figure 10) and the corresponding Fano resonance modulation is shown in **Figure 2d**. Each of the FRs were switched through varied multiple states to the final complete Fano resonance modulation. It is important to note that to achieve spatial selectivity, we kept the stimulus time relatively short (~15 s) and increased the current value, to avoid the spread of thermal energy to other FRs in the SLM. The maximum switching current required for complete modulation of Fano resonances for all FRs were between 650 - 850 mA. The discrepancy in the critical value of the Fano resonance switch-off current can be avoided through improved design for electrical signal routing and thermal energy transfer in the system. Since the current induced changes are fundamentally thermal driven, the switching states are non-volatile. Hence, electrically control enabled spatial selectivity with multilevel non-volatile resonance switching of terahertz waves is clearly demonstrated.

The earlier proposed switching schemes of using thermal and electrical stimulus provide slower switching speeds in timescales of seconds or higher, which is desirable for certain applications.

Alternatively, ultrafast control of terahertz waves is critical for development of modulators, encoders and multiplexers to be used in high-speed terahertz wireless communication systems.[43, 44] Here, we exploit the semiconducting nature of GST in the different crystallographic phases to achieve variable ultrafast volatile terahertz switching under optical excitation. The band gap of amorphous and crystalline GST is reported to be ~ 0.8 eV and 0.5 eV, respectively.[39] When the GST film is optically pumped with photon energy of 1.55 eV, the free carriers generated increases the film conductivity, and hence leads to a corresponding decrease in the terahertz transmission. These photoexcited free carriers decay rapidly to their equilibrium state in picosecond timescales. The precise recombination rate which determines the switching speed of the device, can be manipulated by varying the crystallinity of GST film. The ease of controlling the crystalline order of chalcogenide phase change materials provides an important advantage over other semiconducting materials such as silicon,[13] germanium,[11] gallium arsenide,[3] perovskites,[12] and 2D materials[17]. The photoexcitation and relaxation dynamics of charge carriers in as-deposited, 180 °C and 260 °C annealed GST film with 1.55 eV optical pump is measured using the time resolved terahertz time domain spectroscopy system (see Methods). The normalized differential change in terahertz transmission due to pump ($|\Delta T/T|$) with varying pump-probe delay time ($\Delta \tau$) for the fluence of 636.6 μJ/cm$^2$ is shown in **Figure 3a**. The charge carrier dynamics of GST thin film were also measured for pump fluences of 127.3, 318.3, 636.6, 954.9, 1273.2 μJ/cm$^2$ (see Supplementary Figure 11). The photoconductivity was extracted from maximum value of $|\Delta T/T|$ as shown in **Figure 3b** (see Methods and Supplementary Figure 12)**.** For a given pump fluence, the photoconductivity was lower for as-deposited case and increases strongly for 180 °C and 260 °C annealed GST films. However, the relaxation process was much faster for as-deposited case compared to 180 °C and 260 °C annealed GST film as seen from the extracted decay constant shown in **Figure 3c** (see Supplementary Figure 13). Furthermore, the charge carrier dynamics of 180 °C annealed GST thin film with varying annealing time was characterized (see Supplementary Figure 14). Interestingly, the charge carrier dynamics of the GST film annealed for even 5 mins was similar to the 60 mins annealed film. From application perspective, the choice of crystalline order in GST film would be determined based on the desired switching power and operational speed of the devices, which form the trade-off parameter set.

To determine the device level performance, the terahertz transmission response of as-deposited GFMD was characterized with varying optical pump fluence as schematically shown in **Figure 3d** (see Supplementary Figure 15a) and the corresponding Fano resonance modulation is shown in **Figure 3e**. The Fano resonance modulation increases with pump fluence due to the increase in photoconductivity of as-deposited GST film, however it saturates at ~500 $\mu J/cm^2$, corresponding to the photoconductivity of ~15 S/cm. To probe the optical response of crystalline GST, the annealing time at 180 °C for as-deposited GFMD was limited to 10 mins, since 60 mins annealing would completely quench the Fano resonance as shown **Figure 1e**. The Fano resonance was modulated due to increased terahertz conductivity of the GST film, but was still experimentally observable. More importantly, the ultrafast optical response of the GST film annealed for 10 mins was similar to 60 mins annealed films (see Supplementary Figure 14). The terahertz transmission response of 180 °C annealed GFMD with varying pump fluences was measured (see Supplementary Figure 15b) and the corresponding Fano resonance modulation is shown in **Figure 3e**. The critical pump fluence required for complete modulation of Fano resonance for 180 °C annealed GFMD was ~200 $\mu J/cm^2$, which is less than half as that of as-deposited case (~500 $\mu J/cm^2$). Hence, GST provides a means of not only achieving a non-volatile modulation based on the base conductivity through thermal annealing, but also an ultrafast volatile modulation over the base modulation through optical pumping. More precisely, the Fano modulation in both as-deposited and 180 °C annealed GFMDs could be recovered in picosecond timescales, which is evident from the charge carrier dynamics shown in **Figure 3a**. The ultrafast recovery of Fano resonance in the as-deposited GFMD was experimentally probed by varying the optical pump - terahertz probe delay time (Δτ) for the pump fluence of 636.6 μJ/cm$^2$ as shown in **Figure 3f**. When the optical pump and terahertz probes are time matched (Δτ = 0 ps), the Fano resonance starts to modulate and at the maximum |ΔT/T| (Δτ ~4 ps), the resonance is completely modulated. As the delay is further increased (Δτ ~6 ps), the Fano resonance remains modulated until the critical number of carriers relax back and then the resonance starts to recover. Complete recovery of Fano resonance was observed at ~19 ps. The possibility to manipulate this ultrafast feature by altering the phase composition of GST would provide all-optical ultrafast terahertz modulators with variable modulation depth and switching speeds for the high-speed communication channels.

To showcase the versatility of GST for terahertz manipulation, we fabricated the GFMD on flexible substrates as shown in **Figure 4a**. Conventionally, reconfigurable metamaterials based on phase change materials have been limited to rigid substrates, owing to the need for lattice matching or higher temperature processing.[22, 24-26] The ease of depositing GST thin film using standard RF sputtering process allowed for the development of multifunctional terahertz flexible metadevices. We utilize polyimide as the substrate material, due to its high mechanical flexibility, low terahertz absorption and high thermal stability. The measured terahertz transmission response of Fano resonator array without GST, with as-deposited, 180 °C and 260 °C annealed GST shown in **Figure 4b** is identical to the results observed in GFMDs fabricated on rigid quartz substrate shown in **Figure 1d**. The observed blue shift of the resonances in the flexible GFMD is due to the lower refractive index of polyimide (n ~ 1.72) compared to quartz (n ~ 1.95) at terahertz frequencies. We could even replicate the exponential dependence of Fano modulation with respect to annealing time which shows the good quality and repeatability of the fabricated flexible GFMDs (see Supplementary Figure 16). The robustness of the fabricated flexible GFMD to curvature was characterized by bending the samples along the X-and Y-direction (see Supplementary Figure 17). Curving the GFMD along X as central axis causes the intercellular distance between the Fano resonators to increase. Since the periodicity of the planar Fano resonator was designed to minimize nearest-neighbour coupling, any increase in intercellular distance will not affect the metamaterial response significantly. Hence, GFMD shows robust performance for bending along X-direction up to the curvature of 1.1 cm$^{-1}$ as shown in **Figure 4c**. However, at much higher curvatures, the number of Fano resonators excited will be drastically reduced and hence, the metamaterial response which comes from the collective excitation of resonators is weakened. On the other hand, when GFMD is bent along Y as central axis, the coupling distance between the two dipoles within the unit cell (i.e. split capacitive gaps of Fano resonator) increases. As capacitive gaps in Fano resonators are extremely sensitive, any changes in this region will lead to a drastic change in the metamaterial resonance. Furthermore, the change in the capacitive gaps of the Fano resonator varies from the centre to edge of the sample, guided by the bending curvature. Hence, the metamaterial response is strongly modulated, even for small curvatures as shown in **Figure 4c**. This limitation of the flexible GFMD device comes from the ultrahigh sensitivity of Fano resonance but can be readily overcome by using alternate resonator designs.

GST as a phase change material system provides unique response to varied external stimuli such as thermal, electrical and optical, and combined with the functionally rich metamaterial platform would be a route for developing novel and advanced manipulation of terahertz waves. The multilevel non-volatile resonance switching of thermally triggered GFMDs provides a pathway for memory metadevices,[15, 22] OTP logic devices,[45] variable attenuators, and frequency agile filters operating in terahertz spectral region. Furthermore, based on the added spatially selective control enabled by electrical stimulus, we can envision novel photonic devices such as neuromorphic metadevices, where the single final state of the system will be determined by the combined response of the electrically pre-weighted unit-cells of the GFMD to the incoming terahertz signal.[46] The optical response of GST provides a completely new palette of material properties such as ultrafast response in picosecond timescales, volatile resonance switching, and non-contact terahertz control. More importantly, the terahertz properties of GST can be altered continuously by varying its phase composition. The possibility of using these properties of GST in conjunction would enable a complete control of terahertz waves in terms of spectral, spatial and temporal domains. The ease of translation of GST into flexible devices expands the potential application scope of functionally rich terahertz metadevices involving irregular geometries and topologies. However currently, large area reamorphization of crystalline phases of GST is a challenge that needs to be overcome for the potential commercialization of GST based terahertz metadevices. The earlier reports on GST based reconfigurable photonic devices have been limited to nanoscale or microscale dimensions and multicycle phase reconfiguration of GST over an area of few 100's of $\mu m^2$ is well-established using ultrafast laser pulse,[36] but scaling it to $cm^2$ is a technical challenge. Recently, reamorphization of GST thin film over large area has been successfully demonstrated using irradiation with Ar+[47] and Ge+[48] ions. This could provide a viable pathway for the realization of reversible and multicycle reconfigurable GST based large scale terahertz metadevices.

In summary, the chalcogenide phase change material germanium antimony telluride with its varied and unique material properties and functionally diverse metamaterial forms perfectly complementing technologies for multidimensional control of terahertz waves. The proposed hybrid GST integrated Fano resonant metadevices show novel terahertz functionalities such thermally triggered analog switching states with non-volatility, temporally and spatially resolved resonance switching with electrical stimulus and variable ultrafast, volatile resonance switching with optical control. The most important advantage provided by GST is that all of these unique responses for different external stimulus can be operated in tandem to achieve functionally advanced terahertz devices. Additionally, the ease of translation of these highly functional metadevices to flexible substrates opens up its wide application potential involving irregular curvatures and random topologies. Thus, our hybrid phase change–photonic framework is highly functional, extremely versatile and spectrally scalable and will aid in the development of a wide array of novel high-performance terahertz metadevices.

**Methods**

**<u>Sample Preparation:</u>** Single crystal quartz substrates were cleaned using acetone, IPA and water in ultrasonicator. 200 nm thick GST film was RF sputter deposited with chamber base pressure of 5 x $10^{-6}$ mTorr and RF power of 5W. Photolithography was carried out to define the Fano resonator array with critical dimension of 3 μm over the area of 1 cm x 1 cm. 200 nm thick aluminium was thermally evaporated with base pressure of 7 x $10^{-6}$ mTorr. Lift-off was carried out in acetone to fabricate the Al Fano resonators on GST thin films. Similar fabrication process was used for the fabrication of GFMD in polyimide substrate.

**<u>Time resolved terahertz time domain spectroscopy (TR-TDS):</u>** The TRTS measurements were carried out using Optical-pump-Terahertz-probe setup that is based on ZnTe nonlinear terahertz generation and detection. Optical laser beams used for the generation of terahertz and for photoexcitation of the GST sample are derived from the pulsed beam of energy 6 mJ/pulse that has a pulse width of ~ 35 fs and repetition rate of 1 kHz. The photo-excitation pulse has higher photon energy (1.55 eV) than the band gap of the GST thin film, which is about 0.5 – 0.8 eV (for all phases). The optical pump beam has a beam diameter of approximately 10 mm, which is larger than the focused terahertz beam diameter of nearly 4 mm at the sample position, providing uniform photoexcitation over the GST thin film. The time delay (Δτ) between optical-pump and terahertz probe pulses was controlled by using a translational stage and the pump time delay for the terahertz modulation measurements in the GFMD sample is set at the position where the photoexcited signal is the maximum ($\Delta\tau \sim 4$ ps). At this maximum pump signal, the terahertz scan was performed on the GFMD sample and the reference substrate. The electric field of the terahertz waves was polarized perpendicular to the gap bearing side of the ASR resonators and was incident normally on to the GFMD surface. Later, in the post processing steps the spectrum through the sample ($E_S(\omega)$) is normalized to the reference substrate transmission spectra ($E_R(\omega)$) using the relation $|T(\omega)| = |E_S(\omega)|/|E_R(\omega)|$.

## Acknowledgments

The authors acknowledge the research funding support from National Research Foundation Singapore and Agence Nationale de la Recherche, France- NRF2016- ANR004 (M4197003) and NRF CRP on Oxide Electronics on silicon Beyond Moore (NRF-CRP15-2015-01).

## Additional Information

**Competing financial interests:** The authors declare no competing financial interests.

## Author contributions

P.P. and R.S. conceived and designed research. P.P. and S.P. fabricated the samples. P.P. and A.K. performed the terahertz characterization of the samples. H.J. and S.P. performed the Raman spectroscopy and X-ray diffraction measurements, respectively. P.P. and R.S. wrote the manuscript based on the input of all authors. R.S. and T.V. supervised the project. All authors analysed the data and discussed the results.

**Data Availability**

The data that support the findings of this study are available from the corresponding author on reasonable request.

**References:**

1. M. Tonouchi, Nature Photonics **1**, 97 (2007).
2. A. Jenkins, Nature Photonics (2006).
3. W. J. Padilla, A. J. Taylor, C. Highstrete, M. Lee and R. D. Averitt, Physical review letters **96** (10), 107401 (2006).
4. H.-T. Chen, W. J. Padilla, J. M. O. Zide, A. C. Gossard, A. J. Taylor and R. D. Averitt, Nature **444** (7119), 597-600 (2006).
5. H.-T. Chen, J. F. O'hara, A. K. Azad, A. J. Taylor, R. D. Averitt, D. B. Shrekenhamer and W. J. Padilla, Nature Photonics **2** (5), 295-298 (2008).
6. K. Fan and W. J. Padilla, Materials Today **18** (1), 39-50 (2015).
7. R. Degl'Innocenti, J. Kindness Stephen, E. Beere Harvey and A. Ritchie David, in *Nanophotonics* (2018), Vol. 7, pp. 127.
8. L. Cong, N. Xu, J. Han, W. Zhang and R. Singh, Adv Mater **27** (42), 6630-6636 (2015).
9. L. Liang, M. Qi, J. Yang, X. Shen, J. Zhai, W. Xu, B. Jin, W. Liu, Y. Feng, C. Zhang, H. Lu, H.-T. Chen, L. Kang, W. Xu, J. Chen, T. J. Cui, P. Wu and S. Liu, Advanced Optical Materials **3** (10), 1311-1311 (2015).
10. J. F. O'Hara, R. Singh, I. Brener, E. Smirnova, J. Han, A. J. Taylor and W. Zhang, Optics Express **16** (3), 1786-1795 (2008).
11. X. Lim Wen, M. Manjappa, K. Srivastava Yogesh, L. Cong, A. Kumar, F. MacDonald Kevin and R. Singh, Advanced Materials **30** (9), 1705331 (2018).
12. M. Manjappa, Y. K. Srivastava, A. Solanki, A. Kumar, T. C. Sum and R. Singh, Advanced Materials **29** (32) (2017).
13. M. Manjappa, Y. K. Srivastava, L. Cong, I. Al-Naib and R. Singh, Adv Mater **29** (3) (2017).
14. S. Savo, D. Shrekenhamer and J. Padilla Willie, Advanced Optical Materials **2** (3), 275-279 (2014).
15. W. Y. Kim, H.-D. Kim, T.-T. Kim, H.-S. Park, K. Lee, H. J. Choi, S. H. Lee, J. Son, N. Park and B. Min, Nature Communications **7**, 10429 (2016).
16. L. Ju, B. Geng, J. Horng, C. Girit, M. Martin, Z. Hao, H. A. Bechtel, X. Liang, A. Zettl, Y. R. Shen and F. Wang, Nature Nanotechnology **6**, 630 (2011).
17. Y. K. Srivastava, A. Chaturvedi, M. Manjappa, A. Kumar, G. Dayal, C. Kloc and R. Singh, Advanced Optical Materials **5** (23), 1700762 (2017).
18. R. Singh, J. Xiong, A. K. Azad, H. Yang, S. A. Trugman, Q. Jia, A. J. Taylor and H.-T. Chen, Nanophotonics **1** (1), 117-123 (2012).
19. B. Jin, C. Zhang, S. Engelbrecht, A. Pimenov, J. Wu, Q. Xu, C. Cao, J. Chen, W. Xu and L. Kang, Optics express **18** (16), 17504-17509 (2010).

20. J. Gu, R. Singh, Z. Tian, W. Cao, Q. Xing, M. He, J. W. Zhang, J. Han, H.-T. Chen and W. Zhang, Applied Physics Letters **97** (7) (2010).
21. Y. K. Srivastava, M. Manjappa, L. Cong, H. N. S. Krishnamoorthy, V. Savinov, P. Pitchappa and R. Singh, Advanced Materials **30** (29), 1801257 (2018).
22. T. Driscoll, H.-T. Kim, B.-G. Chae, B.-J. Kim, Y.-W. Lee, N. M. Jokerst, S. Palit, D. R. Smith, M. Di Ventra and D. N. Basov, Science **325** (5947), 1518-1521 (2009).
23. M. Liu, H. Y. Hwang, H. Tao, A. C. Strikwerda, K. Fan, G. R. Keiser, A. J. Sternbach, K. G. West, S. Kittiwatanakul, J. Lu, S. A. Wolf, F. G. Omenetto, X. Zhang, K. A. Nelson and R. D. Averitt, Nature **487**, 345 (2012).
24. Y. Zhao, Y. Zhang, Q. Shi, S. Liang, W. Huang, W. Kou and Z. Yang, ACS Photonics **5** (8), 3040-3050 (2018).
25. C. H. Kodama and R. A. C. Jr., Applied Physics Letters **108** (23), 231901 (2016).
26. H. Cai, S. Chen, C. Zou, Q. Huang, Y. Liu, X. Hu, Z. Fu, Y. Zhao, H. He and Y. Lu, Advanced Optical Materials **6** (14), 1800257 (2018).
27. M. Seo, J. Kyoung, H. Park, S. Koo, H.-s. Kim, H. Bernien, B. J. Kim, J. H. Choe, Y. H. Ahn, H.-T. Kim, N. Park, Q. H. Park, K. Ahn and D.-s. Kim, Nano Letters **10** (6), 2064-2068 (2010).
28. D. Wang, L. Zhang, Y. Gu, M. Q. Mehmood, Y. Gong, A. Srivastava, L. Jian, T. Venkatesan, C.-W. Qiu and M. Hong, Scientific Reports **5**, 15020 (2015).
29. T. H. Jeong, M. R. Kim, H. Seo, S. J. Kim and S. Y. Kim, Journal of Applied Physics **86** (2), 774-778 (1999).
30. M. Wuttig, H. Bhaskaran and T. Taubner, Nature Photonics **11**, 465 (2017).
31. M. Wuttig, Nature Materials **4**, 265 (2005).
32. C. Ríos, M. Stegmaier, P. Hosseini, D. Wang, T. Scherer, C. D. Wright, H. Bhaskaran and W. H. P. Pernice, Nature Photonics **9**, 725 (2015).
33. P. Hosseini, C. D. Wright and H. Bhaskaran, Nature **511**, 206 (2014).
34. C. Ríos, P. Hosseini, R. A. Taylor and H. Bhaskaran, Advanced Materials **28** (23), 4720-4726 (2016).
35. B. Gholipour, J. Zhang, K. F. MacDonald, D. W. Hewak and N. I. Zheludev, Adv Mater **25** (22), 3050-3054 (2013).
36. Q. Wang, E. T. F. Rogers, B. Gholipour, C.-M. Wang, G. Yuan, J. Teng and N. I. Zheludev, Nature Photonics **10**, 60 (2015).
37. N. Raeis-Hosseini and J. Rho, Materials **10** (9) (2017).
38. H. K. Peng, K. Cil, A. Gokirmak, G. Bakan, Y. Zhu, C. S. Lai, C. H. Lam and H. Silva, Thin Solid Films **520** (7), 2976-2978 (2012).
39. B.-S. Lee, J. R. Abelson, S. G. Bishop, D.-H. Kang, B.-k. Cheong and K.-B. Kim, Journal of Applied Physics **97** (9), 093509 (2005).
40. M. Wuttig and N. Yamada, Nature Materials **6**, 824 (2007).
41. W. L. Chan, H.-T. Chen, A. J. Taylor, I. Brener, M. J. Cich and D. M. Mittleman, Applied Physics Letters **94** (21), 213511 (2009).
42. P. Pitchappa, M. Manjappa, C. P. Ho, R. Singh, N. Singh and C. Lee, Advanced Optical Materials **4** (4), 541-547 (2016).
43. T. Nagatsuma, G. Ducournau and C. C. Renaud, Nature Photonics **10**, 371 (2016).
44. T. Nagatsuma, S. Horiguchi, Y. Minamikata, Y. Yoshimizu, S. Hisatake, S. Kuwano, N. Yoshimoto, J. Terada and H. Takahashi, Optics Express **21** (20), 23736-23747 (2013).
45. M. Manjappa, P. Pitchappa, N. Singh, N. Wang, N. I. Zheludev, C. Lee and R. Singh, Nature Communications **9** (1), 4056 (2018).
46. X. Lin, Y. Rivenson, N. T. Yardimci, M. Veli, Y. Luo, M. Jarrahi and A. Ozcan, Science (2018).
47. R. De Bastiani, A. M. Piro, M. G. Grimaldi and E. Rimini, Nuclear Instruments and Methods in Physics Research Section B: Beam Interactions with Materials and Atoms **257** (1), 572-576 (2007).
48. R. D. Bastiani, E. Carria, S. Gibilisco, A. Mio, C. Bongiorno, F. Piccinelli, M. Bettinelli, A. R. Pennisi, M. G. Grimaldi and E. Rimini, Journal of Applied Physics **107** (11), 113521 (2010).

**Figures**

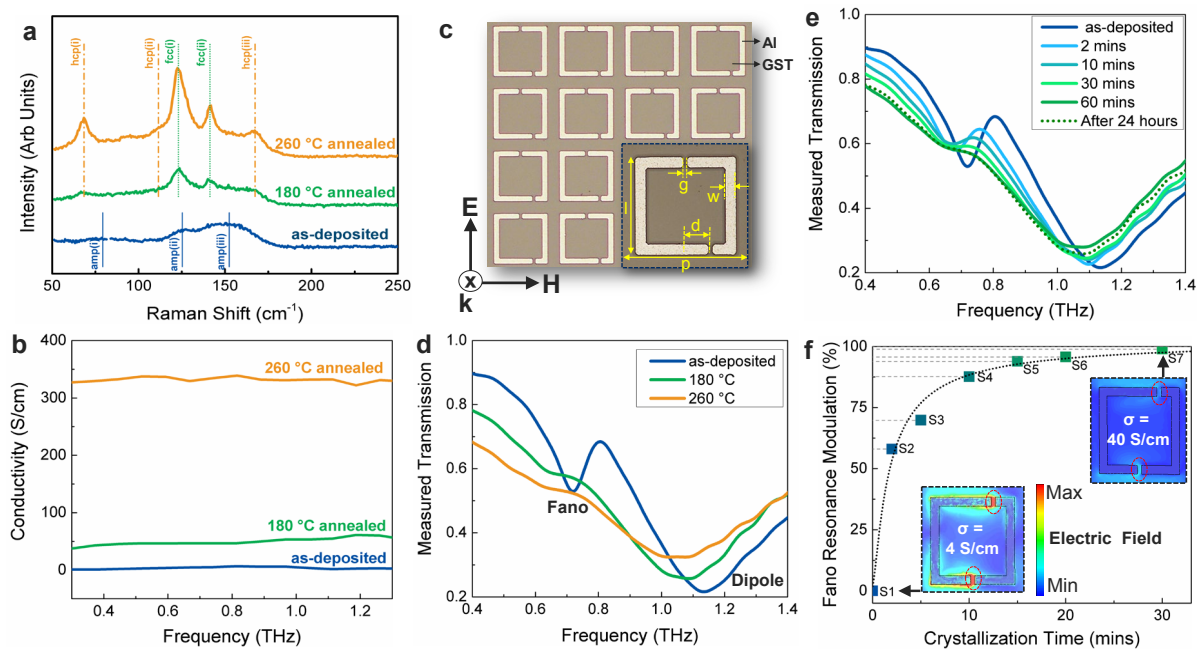

**Figure 1: Thermally switched GFMD with multilevel non-volatile response.** (a) Measured Raman spectra of GST thin film - as-deposited, 180 °C and 260 °C annealed for 60 mins. The Raman spectra clearly shows the evolution of crystallization peaks with increased annealing temperature. (b) Extracted terahertz conductivity of as-deposited, 180 °C and 260 °C annealed GST thin film. (c) Fabricated GFMD with inset showing the geometrical parameters: p = 75 μm, l = 60 μm, w = 6 μm, g = 3 μm and d = 15 μm, respectively. (d) Measured terahertz transmission response of GFMD with as-deposited, 180 °C and 260 °C annealed GST. The as-deposited case shows strong excitation of Fano and dipole resonances. When annealed at 180 °C for 60 mins, the Fano resonance is completely modulated while an appreciable modulation of dipole resonance is observed. With further annealing to 260 °C for 60 mins, much stronger modulation of dipole resonance is observed. (e) Measured multilevel Fano resonance modulation of GFMD by controlling the annealing time at 180 °C. The dotted line shows the thermally switched GFMD after 24 hours thereby confirming the non-volatility of the samples with long retention time and zero hold power (f) Calculated Fano resonance modulation with respect to annealing time at 180 °C and the dashed line shows a single exponential fit. The bottom inset shows the strong field confinement when conductivity of GST is 4 S/cm (as-deposited case), while the top inset shows that the field confinement is completely quenched for 40 S/cm (180 °C annealed for 60 mins case).

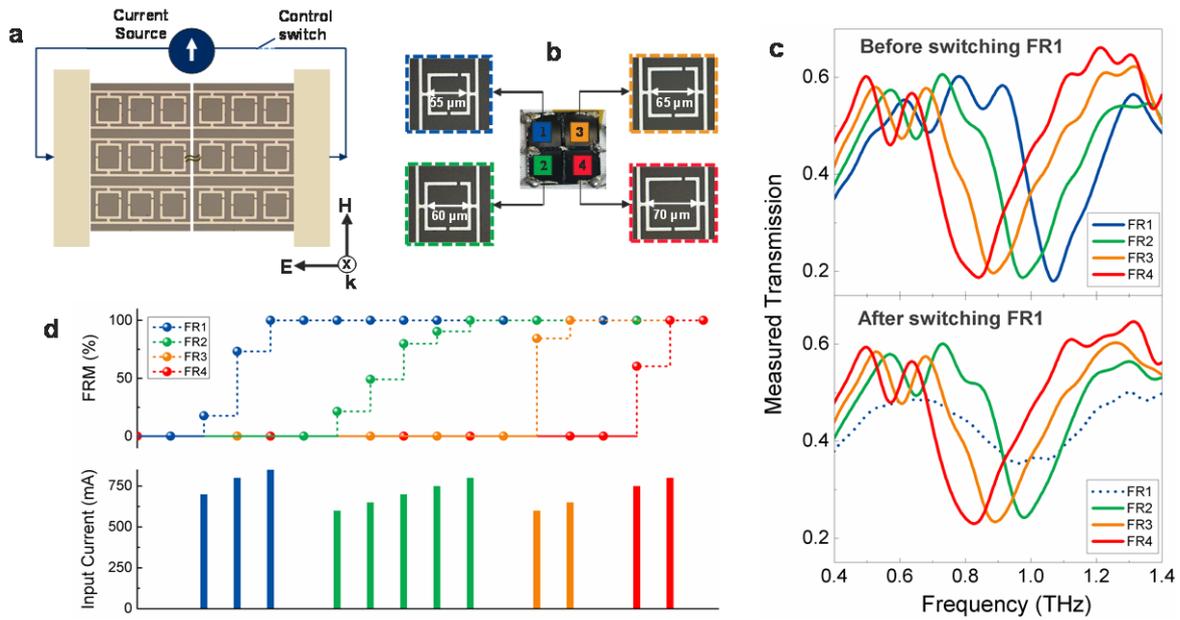

**Figure 2: Electrically switched GFMD with spatially selective non-volatile response.** (a) Schematics of electrically switched GFMD with all unit cells electrically connected across two bond pads. (b) Fabricated SLM sample with four electrically isolated quadrants with insets showing the corresponding unit cells: FR1 – 55 μm, FR2 – 60 μm, FR3 – 65 μm and FR4 – 70 μm, respectively. (c) Measured terahertz transmission response of all FRs before (top graph) and after (bottom graph) switching of FR1 by gradually increasing the current from 0 mA to 850 mA with hold time of 15 s at each current value. This shows the selective switching of FR1, while the other FRs in the SLM remains relatively unchanged, thereby highlighting the spatial light modulation feature. (d) Experimental demonstration of sequential switching of Fano resonance of FR1, FR2, FR3 and FR4 through multiple states by selectively biasing the desired part of the SLM.

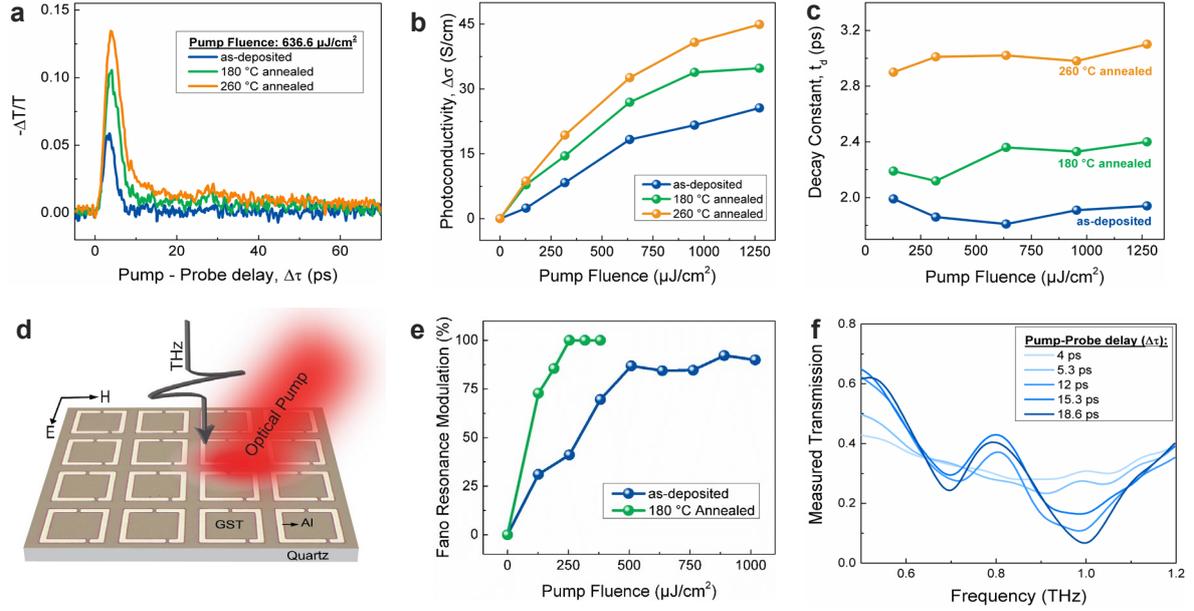

**Figure 3: Optically triggered variable ultrafast resonance modulation.** (a) Measured charge carrier dynamics of optically pumped (1.55 eV) GST thin film in as-deposited, 180 °C and 260 °C annealed at the pump fluence of 636.6 μm/cm$^2$. The |ΔT/T| value and relaxation time constants increases with increased crystalline order in the GST thin film. Extracted values of (b) photoconductivity and (c) decay constant for as-deposited, 180 °C and 260 °C annealed GST thin films for varying pump fluences. (d) Schematics of the optically controlled terahertz response of GFMDs. (d) Measured Fano resonance modulation of optically pumped as-deposited and 180 °C annealed GFMDs. The critical power required for switching off the Fano resonance was almost twice for as-deposited case compared to the 180 °C annealed case. (f) Ultrafast resonance recovery observed form measured terahertz transmission response of as-deposited GFMD with varying pump-probe delay time (Δτ) for pump fluence of 636.6 μm/cm$^2$. The Fano resonance is completely modulated at time matched condition (Δτ = 4 ps), when |ΔT/T| is maximum and remains modulated until Δτ = 6 ps. The Fano resonance recovers to 35 % at 12 ps and complete recovery is achieved at ~19 ps.

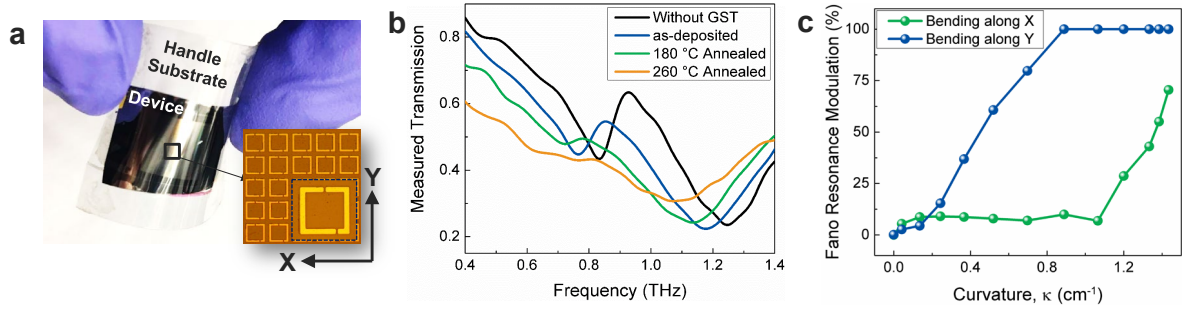

**Figure 4: GFMD on flexible substrate.** (a) Optical image of the fabricated GFMD on flexible polyimide substrate, attached to PET sheet with a central hole as handle substrate for the measurements. The inset shows the optical image of the fabricated flexible GFMD and its unit cell. (b) Measured terahertz transmission response of the fabricated flexible GFMD – without GST, as-deposited GST, and after annealing for 180 °C and 260 °C for 60 mins. The inclusion of as-deposited GST shows a spectral shift and modulation of the Fano resonance. The spectral shift and resonance modulation increased with further annealing at higher temperatures. (c) Measured terahertz transmission spectra of flexible GFMDs under varying curvature along X- and Y-direction. Bending along X- central axis shows no change in Fano resonance strength until 1.1 cm$^{-1}$, however for bending with Y-central axis shows strong modulation of Fano resonance even from 0.1 cm$^{-1}$.